\documentclass[11pt]{article}
\pdfoutput=1
\usepackage{jcappub,amsmath,amssymb}
\usepackage{bm}
\usepackage[table]{xcolor}
\usepackage{graphics}
\usepackage{tikz}
\usetikzlibrary{arrows,positioning,decorations.markings,decorations.pathmorphing,calc}

\definecolor{hyperref}{RGB}{026,028,087}

\def\gsim{ \lower .75ex \hbox{$\sim$} \llap{\raise .27ex \hbox{$>$}} }
\def\lsim{ \lower .75ex \hbox{$\sim$} \llap{\raise .27ex \hbox{$<$}} }
\def\be{\begin{equation}}
\def\ee{\end{equation}}
\def\bea{\begin{eqnarray}}
\def\eea{\end{eqnarray}}

\newcommand{\ba}{\begin{array}}
\newcommand{\ea}{\end{array}}
\newcommand{\mn}{\mu\nu}

\newcommand{\commentout}[1]{}

\newcommand{\cS}{{\cal{S}}}
\newcommand{\cI}{{\cal{I}}}

\newcommand{\comment}[1]{}

\newcommand{\bs}{\begin{split}}

\def\ba{\begin{eqnarray}}
\def\ea{\end{eqnarray}}

\def\nn{\nonumber}

\def\({\left(}
\def\){\right)}

\definecolor{jn}{RGB}{10, 10, 200} 
\definecolor{js}{RGB}{204, 0, 0} 
\definecolor{pgf}{RGB}{10, 150, 10} 

\newcommand*{\mathcolor}{}
\def\mathcolor#1#{\mathcoloraux{#1}}
\newcommand*{\mathcoloraux}[3]{%
  \protect\leavevmode
  \begingroup
    \color#1{#2}#3%
  \endgroup
}
\newlength{\stheight}
\newcommand\textst[1][fu-grey]{
	\ifmmode\setlength{\stheight}{+1.0ex}
	\else\setlength{\stheight}{+0.5ex}
	\fi
	\bgroup\markoverwith{\textcolor{#1}{\rule[\the\stheight]{2pt}{1.0pt}}}\ULon
} 

\newcommand{\textins}[2][fu-grey]{
	\ifmmode\mathcolor{#1}{#2}
	\else\textcolor{#1}{#2}\@\,
	\fi
}

\newcommand{\w}{\wedge}

\newcommand{\N}{{\cal N}}

\newcommand{\EI}{{E_{^{\!\hspace{.2mm}(i)}}\!}}

\newcommand{\LI}{{\Lambda_{^{\!\hspace{.2mm}(i)}}\!}}

\newcommand{\fI}{{f_{^{\!\hspace{.2mm}(i)}}\!}}

\def\({\left(}
\def\){\right)}

\graphicspath{{./}}
\notoc

  \tikzstyle{vecArrow} = [thick, decoration={markings,mark=at position
   1 with {\arrow[semithick]{open triangle 60}}},
   double distance=1.4pt, shorten >= 5.5pt,
   preaction = {decorate},
   postaction = {draw,line width=1.4pt, white,shorten >= 4.5pt}]

\begin{document}

\title{The coupling to matter in \\ Massive, Bi- and Multi-Gravity}


\author[a,b]{Johannes Noller}
\author[b]{, Scott Melville}

\affiliation[a]{Astrophysics, University of Oxford, DWB, Keble Road, Oxford, OX1 3RH, UK} 
\affiliation[b]{The Queen's College, High Street, Oxford, OX1 4AW, UK}

\emailAdd{noller@physics.ox.ac.uk}
\emailAdd{scott.melville@queens.ox.ac.uk}

\abstract{In this paper we construct a family of ways in which matter can couple to one or more `metrics'/spin-2 fields in the vielbein formulation. We do so subject to requiring the weak equivalence principle and the absence of ghosts from pure spin-2 interactions generated by the matter action. Results are presented for Massive, Bi- and Multi-Gravity theories and we give explicit expressions for the effective matter metric in all of these cases. 
}


\maketitle
\newpage


\section{Introduction} \label{sec-intro}

Our understanding of theories of interacting spin-2 fields has experienced significant progress in the past few years. In a development mimicking the way in which interacting spin-1 theories were discovered and explored in the past -- leading from massless Maxwell-like and Yang-Mills theories all the way to the construction of the bosonic sector of the Standard Model -- consistent theories for the spin-2 sector have been identified. Taking general relativity (the theory of a massless spin-2 field) as a starting point, consistent theories of a single massive spin-2, ghost-free (dRGT) Massive Gravity \cite{deRham:2010ik,deRham:2010kj,Hassan:2011hr}, were discovered recently and have since been extended to interacting theories of two spin-2 fields, Bigravity \cite{Hassan:2011tf,Hassan:2011zd,Hassan:2011ea}, and Multi-Gravity \cite{Hinterbichler:2012cn}. As a result we have gained significant insight into consistent potential-like self-interactions in the spin-2 sector. For some excellent reviews on the topic see \cite{Hinterbichler:2011tt,deRham:2014zqa}.  

Having constructed such consistent self-interacting spin-2 theories, a crucial question is how these fields couple to matter. This is still a largely unexplored area. A known consistent way of coupling to matter is to simply always minimally couple matter to just one single spin-2 field (this of course breaks any symmetry between multiple spin-2 fields potentially present at the level of their self-interactions). Minimally coupling matter to more than one spin-2 field has been explored in \cite{Tamanini:2013xia,Akrami:2013ffa,Akrami:2014lja}. This, however, breaks the weak equivalence principle and also generically gives rise to the propagation of ghost-like degrees of freedom at unacceptably low energy scales \cite{Yamashita:2014fga,deRham:2014naa}. Here we investigate whether there are any potentially consistent ways of coupling matter to more than one spin-2 field, which still respect the weak equivalence principle. We find that there are and explicitly construct a general family of metrics and vielbeins, which are functions of $\N$ spin-2 fields and which we can use to couple matter to $\N$ spin-2 fields in a way that respects the weak equivalence principle and which does not generically lead to the propagation of any ghost-like degrees of freedom (at least up to the decoupling limit scale $\Lambda_3$ - see \cite{deRham:2014naa}).

While this paper was being finished, \cite{Yamashita:2014fga,deRham:2014naa} appeared, which also investigate aspects of the coupling to matter in Massive and Bigravity. In particular the parts of this paper dealing with Bigravity theories have some overlap with the analysis presented in \cite{deRham:2014naa} (albeit in different formulations -- we work in the vielbein picture, whereas \cite{deRham:2014naa} work in the metric formulation -- and with somewhat different starting points -- our approach constructively generates coupling terms from a classical perspective, whereas \cite{deRham:2014naa} generate their coupling terms by requiring that one-loop effects do not introduce any ghosts). Our approaches are complementary and, where results overlap, they are in agreement.   

Throughout this paper we use the vielbein formulation, which in a gravity context was developed in \cite{Nibbelink:2006sz,Hanada:2008hs,Chamseddine:2011mu,Chamseddine:2011bu,Mirbabayi:2011aa}.  For a thorough discussion of when the metric and vielbein formulations are  equivalent and when they are not, see \cite{Hinterbichler:2012cn,Deffayet:2012zc} - also see \cite{Banados:2013fda,Mourad:2014roa,Deser:2014hga} for constraint analyses of massive gravity models in the vielbein formulation and related outstanding questions e.g. concerning the absence of ghosts in general multi-vielbein theories. This formulation lends itself to a general Multi-Gravity analysis, since interactions between more than two spin-2 fields take on a vastly simpler form than in the metric formulation \cite{Hassan:2012wt}. Moreover, it is straightforward to construct ghost-free Multi-Gravity theories with arbitrary superpositions of interaction terms \cite{Hinterbichler:2012cn,Noller:2013yja} in this picture, especially also in cases where the model's `theory graph' contains closed loops and the equivalence between vielbein and metric pictures breaks down \cite{Hinterbichler:2012cn,Deffayet:2012zc,loops}.
\\

{\it Outline}: The plan for this paper is as follows. In section \ref{sec-MG} we review the construction of consistent theories of $\N$ interacting spin-2 fields -- theories of Massive, Bi- and Multi-Gravity. In section \ref{sec-coupling} we then use those results to generate all candidate couplings to matter of the spin-2 fields subject to the requirements of 1) The weak equivalence principle 2) Ghost-freedom of pure spin-2 interactions generated by the matter action. This constructive argument generates a family of ways in which matter can couple to $\N$ spin-2 fields in $D$ dimensions, where we have assumed that the effective matter metric is a general function of the vielbeins and the Lorentz metric $\eta_{AB}$ only.  We also investigate the cases of 2D Bigravity and 4D Massive and Bigravity in some additional detail. Finally we conclude in section \ref{sec-conc} and have collected a few additional concrete examples illustrating our results in appendix \ref{appendix}.
\\

{\it Conventions}: Throughout this paper we use the following conventions. $D$ refers to the number of spacetime dimensions and we use Greek letters $\mu, \nu, \ldots$ to denote spacetime indices, which are raised and lowered with the metric $g_{\mu\nu}$.  Capital Latin letters $A$, $B,\ldots$ are reserved for Lorentz indices and are raised and lowered with the Minkowski metric $\eta_{AB}$.  Bracketed indices $(i),(j),\ldots$, label the different vielbeins/spin-2 fields -- label indices are not automatically summed over and whether they are upper or lower indices carries no meaning. We denote the completely anti-symmetric epsilon symbol by $\tilde \epsilon$ and define it such that $\tilde\epsilon_{012\cdots D}=1$ regardless of the signature of the metric or the position (up/down) of indices (hence $\tilde\epsilon^{012\cdots D}=\tilde\epsilon_{012\cdots D}=1$).  

\section{Massive, Bi- and Multi-Gravity: Consistently interacting spin-2 fields}\label{sec-MG}

In this section we quickly review the construction of consistent theories of interacting spin-2 fields -- Massive, Bi- and Multi-Gravity -- in the vielbein formulation closely following \cite{Hinterbichler:2012cn}, collecting several useful results for constructing general matter couplings in the next section. 
We begin with an action for $\N$ massless, non-interacting spin-2 fields (or `metrics') $g_{(i)}$
\be \label{S-Nnonint}
\cS = \sum_{i=1}^{\N} {M_{(i)}^{D-2}\over 2} \int d^Dx \sqrt{-g_{(i)}}\,R[g_{(i)}],
\ee
where $R[g_{(i)}]$ is the Ricci scalar formed out of a metric $g_{(i)}$ and having different $M_{(i)}$ allows for different Planck masses for each $g_{(i)}$. We can recast this in a vielbein formulation by defining vielbein one-forms ${\bf E}_{(i)}^A=\EI_\mu^{\ A}(x)dx^\mu$, which satisfy\footnote{Note that the inverse vielbeins $\EI^\mu_{\ A}(x)$ satisfy $\EI^\mu_{\ A}\EI_\mu^{\ B}=\delta_A^B$ and $\EI^\mu_{\ A}\EI_\nu^{\ A}=\delta_\nu^\mu$.}
\be
g_{(i)}{}_{\mu\nu} = \EI_\mu^{\ A} \, \EI_\nu^{\ B}\,\eta_{AB} \,  . 
\label{gEE}
\ee
The action \eqref{S-Nnonint} can then equivalently be written in terms of vielbeins
\be \label{S-Nnonint-vielbein}
\cS = \sum_{i=1}^{\N} {M_{(i)}^{D-2}\over 2} \int d^Dx\ \det(E_{(i)}) R[E_{(i)}],
\ee
since $\sqrt{-g_{(i)}} = \det(E_{(i)})$ and $R^{\mn}_{\alpha\beta} = E_{(i)\alpha}^{\ A} E_{(i)\beta}^{\ B} R^{\mn}_{AB}$, where $R^{\mn}_{AB}$ is the gauge curvature 2-form, so that $R[E_{(i)}] = E_{(i)\mu}^{\ A} E_{(i)\nu}^{\ B} R^{\mn}_{AB} = R[g_{(i)}]$.
Note that, in the absence of further (cross-) interaction terms between the different fields, \eqref{S-Nnonint-vielbein} is invariant under local diffeomorphism symmetries $\fI$ and Lorentz transformations $\LI$ for each field $\EI$ separately 
\be 
\EI_{\mu}^{\ A}(x)\rightarrow {\partial \fI^\nu \over \partial  x^\mu}\EI_{\nu}^{\ A}\left(\fI(x)\right), \, \quad\quad\quad \EI_\mu^{\ A}\rightarrow \LI^{A}_{\ B} \EI^{\ B}_\mu\, .
\ee
\\

{\bf Consistent interaction terms}: What interaction terms between vielbeins can one add to \eqref{S-Nnonint-vielbein} without introducing ghost-like instabilities into the resulting theory? Focussing on non-derivative interactions (most relevant at low energies), consistent Massive Gravity interactions were found in the metric formalism in \cite{deRham:2010ik,deRham:2010kj,Hassan:2011hr} and then generalised to the case of two dynamical spin-2 fields in Bigravity \cite{Hassan:2011tf,Hassan:2011zd,Hassan:2011ea}. In \cite{Hinterbichler:2012cn} these interactions were recast in the vielbein language and generalised to the case of Multi-Gravity with interactions between arbitrarily many spin-2 fields in $D$ dimensions. This led to the conjecture \cite{Hinterbichler:2012cn} that the most general ghost-free interaction term for $\N$ spin-2 fields in $D$ dimensions is\footnote{Regarding a discussion of the `uniqueness' of these interaction terms, at least in the case of Massive Gravity, we refer to \cite{deRham:2010ik,deRham:2010kj,Comelli:2012vz,ASM-thesis,cutoff}.}
\bea \label{genint}
{\hat {\bm \cI}}_{(i_1 i_2 \ldots i_D)} &\equiv & \tilde  \epsilon_{A_1 A_2 \cdots A_D}\, {\bf E}^{A_1}_{(i_1)}\w {\bf E}^{A_2}_{(i_2)}\w \ldots\w {\bf E}^{A_D}_{(i_D)},
\eea
where the indices $(i_1 i_2 \ldots i_D)$ keep track of which fields are interacting and where the wedge product is defined as usual, so that
\bea  
\hat{\bm \cI}_{(i_1 i_2 \ldots i_D)}
 =  \tilde \epsilon_{A_1 A_2 \cdots A_D} \, \tilde  \epsilon^{\mu_1 \mu_2 \cdots \mu_D} \, 
 {E}_{(i_1)}{}_{\mu_1}^{\ A_1} \,  {E}_{(i_2)}{}_{\mu_2}^{\ A_2} \cdots {E}_{(i_D)}{}_{\mu_D}^{\ A_D}  \, d^Dx \equiv {\cI}_{(i_1 i_2 \ldots i_D)} d^Dx\,,
 \label{density}
\eea
where we have defined ${\cI}$ for later convenience. The anti-symmetric nature of the interaction term means the order of labels in \eqref{genint} $(i_1 i_2 \ldots i_D)$ is irrelevant, i.e. $\hat{\bm \cI}_{(i_1 \ldots i_j \ldots i_k \ldots i_D)} = \hat{\bm \cI}_{(i_1 \ldots i_k \ldots i_j \ldots i_D)}$.
Note that, while \eqref{genint} has been conjectured to describe all consistent interaction terms (it certainly includes all the known consistent interactions \cite{deRham:2010ik,deRham:2010kj,Hassan:2011hr,Hassan:2011tf,Hassan:2011zd,Hassan:2011ea,Hinterbichler:2012cn}), it is still an open question whether all solutions of \eqref{genint} (plus Einstein-Hilbert terms for the participating spin-2 fields) are fully ghost-free \cite{Deffayet:2012zc,Banados:2013fda,Mourad:2014roa}. Here, however, we will work with the conjecture that \eqref{genint} does indeed describe all consistent interaction terms and as such we write the general potential for $\N$ spin-2 fields as
\bea
V\left(E_{(i_1)}, \ldots, E_{(i_N)}\right) = \widehat\sum c_{(i_1 i_2 \ldots i_D)} \hat{\bm \cI}_{(i_1 i_2 \ldots i_D)}
\eea
where the $c_{(i_1 i_2 \ldots i_D)}$ are constant coefficients completely symmetric in all $i_j$ and we have defined the ordered sum $\widehat{\sum}$
\be
\widehat{\sum} c_{(i_1 i_2 \ldots i_D)} \hat{\bm \cI}_{(i_1 i_2 \ldots i_D)} \equiv  \sum_{i_1,i_2,\ldots,i_D = 0,1,\ldots, \N}^{i_1 \le i_2 \le \ldots \le i_D} c_{(i_1 i_2 \ldots i_D)} \hat{\bm \cI}_{(i_1 i_2 \ldots i_D)},
\ee
which sums over the indices $(i_1,i_2,\ldots,i_D)$ from 1 to $\N$ for each index $i_j$, subject to the requirement that $i_1 \le i_2 \le \ldots \le i_D$ (otherwise we repeatedly sum over identical terms, since the ordering of the label indices does not matter due to the anti-symmetry of the interaction term). At this point also note that interaction terms break the $\N$ copies of Lorentz and local diffeomorphism synmmetry down to their respective diagonal subgroups.\footnote{This is true as long as the `theory graph' for the model in question is connected. If not, each disconnected `island' will have one remaining copy of Lorentz and local diffeomorphism synmmetry each - see \cite{Hinterbichler:2012cn,Noller:2013yja} for details.} A general Multi-Gravity theory is then governed by the action
\bea 
\label{drgtvielmulti}
\cS &=& \sum_{i} {M_{(i)}^{D-2}\over 2} \int d^Dx\ (\det E_{(i)}) \,R[E_{(i)}] -{m^2 M_{\rm eff}^{D-2}\over 8}\int \widehat\sum c_{(i_1 i_2 \ldots i_D)} \hat{\bm \cI}_{(i_1 i_2 \ldots i_D)}\, ,
\eea
where $M_{\rm eff}$ is the effective Planck mass for the interaction term.\footnote{We may choose any $M_{(i)}$ or some dimensionally correct combination thereof - overall constants of proportionality can be traded between $m,M_{\rm eff}$ and the $c_{(i_1 \ldots i_D)}$'s. For Bigravity the standard choice is $M_{\rm eff}^{D-2} = (1/M_{(1)}^{D-2} + 1/M_{(2)}^{D-2})^{-1}$. Also note that we have assumed standard Einstein-Hilbert derivative interactions in \eqref{drgtvielmulti} - for a discussion of alternative `kinetic terms' see \cite{Folkerts:2011ev,Hinterbichler:2013eza,Kimura:2013ika,deRham:2013tfa}.}
\\

{\bf Interaction terms in 4D}: As a concrete example let us list all possible non-derivative interaction terms in 4D described by \eqref{genint}. For a single spin-2 field and hence a single vielbein $E_{(i)}{}_{\mu}^{\ A}$ the only term is
\be \label{1in4D}
\cI_{(iiii)} =  \det(E_{(i)}) = \sqrt{-g_{(i)}},
\ee
i.e. a cosmological constant term and the only allowed non-derivative self-interaction term for a spin-2 field in GR (the other allowed term, if we impose second-order equations of motion, is the second Lovelock invariant in 4D, i.e. the derivative Einstein-Hilbert term self-interaction). For two spin-2 fields $E_{(i)}, E_{(j)}$ the possible interactions are
\be \label{2in4D}
 \cI_{(iiii)}, \quad \cI_{(jjjj)}, \quad \cI_{(iiij)}, \quad \cI_{(iijj)}, \quad \cI_{(ijjj)},
\ee
i.e. a copy of the single spin-2 field term for each field and three new cross-interaction terms. The resulting five total terms in fact encapsulate all the interactions of ghost-free Massive and Bigravity \cite{deRham:2010ik,deRham:2010kj,Hassan:2011hr,Hassan:2011tf,Hassan:2011zd,Hassan:2011ea,Hinterbichler:2012cn}. For three spin-2 fields $E_{(i)}, E_{(j)}, E_{(k)}$ we have  the following three-field interaction terms
\be \label{3in4D}
 \cI_{(iijk)}, \quad \cI_{(ijjk)}, \quad \cI_{(ijkk)}.
\ee
Of course in a theory with three fields all the ways of choosing two fields and building interaction terms as in \eqref{2in4D} (9 terms) or building cosmological constant type terms with one field as in \eqref{1in4D} (3 terms) are also allowed.
Finally for four fields $E_{(i)}, E_{(j)}, E_{(k)}, E_{(l)}$ we have all the ways of choosing one, two or three fields and building one-, two- or three-field interaction terms as above and in addition the four-field interaction term
\be \label{4in4D}
 \cI_{(ijkl)},
\ee
resulting in a total of 35 terms for four fields in 4D. Naturally there are no interaction terms for more than $D$ fields in $D$ dimensions of the type considered here, since $\tilde \epsilon$ vanishes if it carries more than $D$ indices.
\\

{\bf $\N$ fields in $D$ dimensions}: In the general case the total number of possible interaction terms (in terms of binomial coefficients $\left(\begin{array}{c}a \\ b\end{array}\right)$) is  
\be 
\sum_n^\N \left(\begin{array}{c}\N \\ n\end{array}\right) \left(\begin{array}{c}D-1 \\ n-1\end{array}\right) = \left(\begin{array}{c}\N+D-1 \\D\end{array}\right),
\ee
where the first binomial coefficient counts the number of ways of choosing $n$ fields from the $\N$ fields in the theory and the second coefficient counts the number of $n$ field interaction terms in $D$ dimensions.

\section{The coupling to matter} \label{sec-coupling}

Having constructed consistent potential self- and cross-interactions in the spin-2 sector in the previous section, we would now like to know how a theory of one or several interacting spin-2 fields can couple to matter. Here we derive the general form of such a coupling subject to the following conditions. Firstly we require the weak equivalence principle, i.e. the matter coupling in the action will have to take the form
\be \label{matteract}
\int d^D x \sqrt{-\det \left({\bm g}^{(M)}\right)} {\cal L}\left[\Phi_i, {\bm g}^{(M)} \right],
\ee
where $\Phi_i$ labels all the matter fields and all indices in the matter Lagrangian are contracted with the `matter metric' ${\bm g}^{(M)}$. This metric is in principle any function of the $\N$ spin-2 fields -- and hence $\N$ vielbeins $E_{(1)},\ldots,E_{(\N)}$ and their inverses -- we considered above.\footnote{Note that we are solely interested in how matter couples to one or several spin-2 fields here -- to `metrics' if you will -- in the absence of further additional degrees of freedom. If additional scalar or vector fields are introduced there are of course a multitude of interesting and known ways that matter can couple to gravity non-minimally \cite{Copeland:2006wr,Clifton:2011jh}.} Here we will restrict ourselves to matter metrics that are solely functions of the vielbeins $E_{(i)}$ and not of their inverses -- we will complete the picture and discuss the most general matter couplings one can build in the full vielbein picture in \cite{mattercoupling}.
Secondly we restrict the allowed ${\bm g}^{(M)}$ by requiring that the matter coupling term does not introduce ghost-like pure spin-2 interactions. This requirement is a necessary condition for the matter coupling to not introduce a ghost at a scale $\Lambda_g$ below $\Lambda_3$ (which would limit the range of validity of the effective field theory to energy scales up to $\Lambda_g$). More specifically this means we require that the pure spin-2 piece of the matter action, i.e. any contribution from ${\cal L}\left[\Phi_i, {\bm g}^{(M)} \right]$ that looks like a cosmological constant for ${\bm g}^{(M)}$
\be \label{CCon}
\int d^D x \sqrt{-\det \left({\bm g}^{(M)}\right)},
\ee
does not render the theory unstable. We generically expect the `vacuum energy' of the matter fields $\Phi_i$ in \eqref{matteract} to generate such a term (even if there is no bare `cosmological constant' term in ${\cal L}\left[\Phi_i, {\bm g}^{(M)} \right]$). The ${\bm g}^{(M)}$ we are about to construct should therefore be seen as a general way of coupling spin-2 fields to matter in a way that in principle allows the full theory to stay a valid effective field theory at least up to the scale $\Lambda_3$. Depending on the explicit way matter fields $\Phi_i$ enter ${\cal L}\left[\Phi_i, {\bm g}^{(M)} \right]$, the form of the metric may be restricted further given the precise nature of ${\cal L}\left[\Phi_i, {\bm g}^{(M)} \right]$.
\footnote{In other words, given some specific matter Lagrangian ${\cal L}\left[\Phi_i, {\bm g}^{(M)} \right]$ ghost-like degrees of freedom may still appear due to the precise nature of coupling to the $\Phi_i$ at some given energy scale $\Lambda_g$. Below the energy scale $\Lambda_g$ the effective theory would then still be fully consistent and ghost-free. If $\Lambda_g > \Lambda_3$ moreover the theory is completely ghost-free in the decoupling limit and of course a valid effective field theory at least up to $\Lambda_3$. If ghost-freedom is required even above the would-be ghost-scale $\Lambda_g$ this can imply further restrictions on the form of the matter metric ${\bm g}^{(M)}$ -- a possible outcome of this would be that the new matter couplings reduce to simple GR-like minimal couplings. The conditions on the form of the metric derived throughout this paper are therefore {\it necessary} conditions for ghost-freedom. They are also {\it necessary} conditions for the more restricted requirement of ghost-freedom up to $\Lambda_3$. Whether they are sufficient depends on the form of ${\cal L}\left[\Phi_i, {\bm g}^{(M)} \right]$ -- we discuss this below.} 
We will investigate such effects (i.e. ones specific to direct coupling to the $\Phi_i$ matter fields) further in \cite{mattercoupling} -- also see \cite{Yamashita:2014fga,deRham:2014naa} -- but here we will see that the above requirement already severely constrains potential matter metrics ${\bm g}^{(M)}$. For the family of metrics we consider (see below), the restrictions we impose should therefore be seen as {\it necessary} (but depending on the precise form of ${\cal L}\left[\Phi_i, {\bm g}^{(M)} \right]$ not necessarily {\it sufficient}) conditions for ghost freedom. 
\footnote{Naturally the matter Lagrangian ${\cal L}\left[\Phi_i, {\bm g}^{(M)} \right]$ may include ghost-like matter degrees of freedom $\Phi_i$ itself. However, here we solely wish to ensure that the degrees of freedom of the gravity sector do not develop instabilities as a result of the matter coupling. In other words, the matter coupling should not re-introduce e.g. the Boulware-Deser ghost \cite{Boulware:1973my} that the careful construction of spin-2 interaction terms in the previous section ensured is not present prior to adding a matter coupling. An acceptable matter coupling should not render a theory unstable by default.}

Going back to our requirement that \eqref{CCon} -- the cosmological constant term for ${\bm g}^{(M)}$ and a self-interaction term for the $E_{(i)}$ -- does not re-introduce ghost-like degrees of freedom means it must be a superposition of allowed interaction terms \eqref{genint}. As such we require
\be \label{detsolprop}
\sqrt{-\det \left({\bm g}^{(M)}\right)} \propto \widehat\sum c_{(i_1 i_2 \ldots i_D)} {\cI}_{(i_1 i_2 \ldots i_D)}.
\ee
In the following we will therefore find the general solution that satisfies
\be \label{detsol}
-\det \left({\bm g}^{(M)}\right) = \frac{1}{(D!)^2}\left(\widehat\sum c_{(i_1 i_2 \ldots i_D)} {\cI}_{(i_1 i_2 \ldots i_D)}\right)^2
\ee
where we have chosen the constant of proportionality for convenience (it can of course be absorbed into the $c_{(i_1, \ldots,i_D)}$). We will solve \eqref{detsol} to give us the general allowed form of the matter metric. The corresponding solutions for the $c_{(i_1, \ldots,i_D)}$ will determine the form of the vielbein (spin-2) interactions induced by ${\bm g}^{(M)}$.
\\

{\bf The matter metric}: We consider any arbitrary metric that is a rank-2 tensor symmetric in its two (spacetime) indices and which can be built out of the $\N$ vielbeins $E_{(1)},\ldots,E_{(\N)}$ and the (Lorentz) Minkowski metric $\eta_{AB}$.\footnote{We discuss what changes if one considers the most general symmetric objects one can build in the vielbein picture (in particular also using inverse vielbeins as building blocks) as candidates for the matter metric and what consequences this has for the associated constraint analysis in \cite{mattercoupling}.}
\be
{\bm g}^{(M)} = {\bm g}^{(M)}\left[E_{(1)}{}_\mu^{\ A},\ldots,E_{(\N)}{}_\mu^{\ A},\eta_{AB}\right]
\ee
Determining all the possible matter metrics that can be built in this way is now very straightforward. The only type of term that has two free spacetime indices and no free Lorentz indices that can be built with the $E_{(i)}$ and $\eta_{AB}$ is of the form
$E^{(i)}{}_{\mu}^{\  A} E^{(j)}{}_{\nu}^{\ B} \eta_{AB}$,
so that a general matter metric can be written\footnote{We have chosen the form of the constant coefficients in \eqref{genmattermet} -- $\alpha_{(ii)}^2$ for $(ii)$ terms, $\alpha_{(ij)}$ for $(ij)$ terms -- for later convenience. It is important to point out that at this point a negative $\alpha_{(ii)}^2$ is physically allowed. This is because $\alpha_{(ii)}$ never appears by itself, so a negative $\alpha_{(ii)}^2$ does not imply that the matter metric or any other physical quantity is complex. As we will find, we will require $\alpha_{(ii)}^2 \alpha_{(jj)}^2 \ge 0$ for all $i,j$, however.}
\bea
{\bm g}^{(M)}_{\mn} &=& \sum_{i=1}^{\N} \alpha_{(ii)}^2 E^{(i)}{}_{\mu}^{\  A} E^{(i)}{}_{\nu}^{\ B} \eta_{AB} \nn\\
&+& \sum_{i,j=1,\ldots,\N}^{i < j} \alpha_{(ij)} \left( E^{(i)}{}_{\mu}^{\  A} E^{(j)}{}_{\nu}^{\ B} + E^{(j)}{}_{\mu}^{\  A} E^{(i)}{}_{\nu}^{\ B} \right) \eta_{AB},\label{genmattermet}
\eea
where the form of the cross-terms between different $E_{(i)}$ and $E_{(j)}$ in the second line of \eqref{genmattermet} comes from the requirement that ${\bm g}^{(M)}$ is symmetric in its spacetime indices. We will now show by construction what the most general form of \eqref{genmattermet} that leads to an acceptable coupling to matter (i.e. that satisfies \eqref{detsolprop}) is for $\N$ spin-2 fields in D dimensions.

\subsection{A warm-up exercise: Bigravity in 2D}\label{subsec-coup2D}

Let us begin with a suitably simple example -- Bigravity in 2D. This allows us to show explicitly how \eqref{detsol} can be solved in a systematic fashion in order to establish the general permissible form of the matter metric  ${\bm g}^{(M)}$. In the case of Bigravity (regardless of the dimension $D$), \eqref{genmattermet} becomes
\bea
{\bm g}^{(M)}_{\mn} &=& \alpha_{(11)}^2 E_{(1)}{}_{\mu}^{\  A} E_{(1)}{}_{\nu}^{\ B}\eta_{AB} + \alpha_{(22)}^2 E_{(2)}{}_{\mu}^{\  A} E_{(2)}{}_{\nu}^{\ B} \eta_{AB} \nn\\   &+& \alpha_{(12)} \left( E_{(1)}{}_{\mu}^{\  A} E_{(2)}{}_{\nu}^{\ B} + E_{(2)}{}_{\mu}^{\  A} E_{(1)}{}_{\nu}^{\ B}\right) \eta_{AB}.
\label{genmet2D}
\eea
Furthermore, in 2D, we can express $\det \left({\bm g}^{(M)}\right)$ as 
\be \label{det2D}
-\det \left({\bm g}^{(M)}\right) = -\frac{1}{2!}\tilde\epsilon^{\mu_1 \mu_2}\tilde\epsilon^{\nu_1 \nu_2} {\bm g}^{(M)}_{\mu_1 \nu_1}{\bm g}^{(M)}_{\mu_2 \nu_2}.
\ee
We would now like to explicitly work this out in terms of the $E_{(i)}$ by substituting \eqref{genmet2D} into \eqref{detsol} (or equivalently by working out \eqref{det2D} using \eqref{genmet2D}). We find 
\bea
-\det \left({\bm g}^{(M)}\right)  &=& \frac{1}{2}\alpha_{(11)}^4 E_{(1)}{}_{\mu A} E_{(1)}{}^{\mu A} E_{(1)}{}_{\nu B} E_{(1)}{}^{\nu B}
-\frac{1}{2} \alpha_{(11)}^4 E_{(1)}{}^{\mu A} E_{(1)}{}_{\mu}^{\ B} E_{(1)}{}^{\nu}_{\ A} E_{(1)}{}_{\nu B} \nn\\
&+& \frac{1}{2} \alpha_{(22)}^4 E_{(2)}{}_{\mu A} E_{(2)}{}^{\mu A} E_{(2)}{}_{\nu B} E_{(2)}{}^{\nu B}
-\frac{1}{2} \alpha_{(22)}^4 E_{(2)}{}^{\mu A} E_{(2)}{}_{\mu}^{\ B} E_{(2)}{}^{\nu}_{\ A} E_{(2)}{}_{\nu B} \nn\\
&+& 2 \alpha_{(11)}^2 \alpha_{(12)} E_{(1)}{}_{\mu A} E_{(1)}{}^{\mu A} E_{(1)}{}^{\nu B} E_{(2)}{}_{\nu B}
-2 \alpha_{(11)}^2 \alpha_{(12)} E_{(1)}{}^{\mu A} E_{(1)}{}_{\mu }^{\ B} E_{(1)}{}^{\nu}_{\ A} E_{(2)}{}_{\nu B} \nn\\
&+& 2 \alpha_{(12)} \alpha_{(22)}^2 E_{(1)}{}^{\mu A } E_{(2)}{}_{\mu A } E_{(2)}{}_{\nu B } E_{(2)}{}^{\nu B }
-2 \alpha_{(12)} \alpha_{(22)}^2 E_{(1)}{}^{\mu A } E_{(2)}{}_{\mu }^{\ B} E_{(2)}{}^{\nu }_{\ A } E_{(2)}{}_{\nu B} \nn\\
&-& \alpha_{(11)}^2 \alpha_{(22)}^2 E_{(1)}{}^{\mu A } E_{(1)}{}^{\nu }_{\ A } E_{(2)}{}_{\mu }^{\ B }  E_{(2)}{}_{\nu B }
+ \alpha_{(11)}^2 \alpha_{(22)}^2 E_{(1)}{}_{\mu A } E_{(1)}{}^{\mu A } E_{(2)}{}_{\nu B } E_{(2)}{}^{\nu B } \nn\\
&-&  \alpha_{(12)}^2 E_{(1)}{}^{\mu A }  E_{(1)}{}^{\nu B } E_{(2)}{}_{\mu B } E_{(2)}{}_{\nu A }
+2 \alpha_{(12)}^2 E_{(1)}{}^{\mu A } E_{(1)}{}^{\nu B } E_{(2)}{}_{\mu A }  E_{(2)}{}_{\nu B } \nn\\
&-&  \alpha_{(12)}^2 E_{(1)}{}^{\mu A } E_{(1)}{}_{\mu }^{\ B} E_{(2)}{}^{\nu }_{\ A } E_{(2)}{}_{\nu B}
\label{LH2D}
\eea
In order to find all solutions that lead to acceptable (ghost-free pure spin-2) vielbein interactions we now need to compute the right-hand side of \eqref{detsol} and find the most general consistent way of matching the coefficients $\alpha_{(ij)}$ and $c_{(ij)}$. Consistent vielbein interactions for Bigravity in 2D are
\bea
\widehat\sum c_{(i_1 i_2)} {\cI}_{(i_1 i_2)} = \tilde\epsilon_{A_1 A_2}\tilde\epsilon^{\mu_1\mu_2} &\Big(& c_{(11)}  E_{(1)}{}_{\mu_1}^{\ A_1} E_{(1)}{}_{\mu_2}^{\ A_2} + c_{(12)} E_{(1)}{}_{\mu_1}^{\ A_1} E_{(2)}{}_{\mu_2}^{\ A_2} \nn\\
&+&c_{(22)} E_{(2)}{}_{\mu_1}^{\ A_1} E_{(2)}{}_{\mu_2}^{\ A_2}\Big),
\label{int-2D}
\eea
so that the right-hand side of \eqref{detsol} becomes
\bea
&& \frac{1}{(D!)^2}\left(\widehat\sum c_{(i_1 i_2)} {\cI}_{(i_1 i_2)}\right)^2 = \nn\\ && \frac{1}{2} c_{(11)}^2 E_{(1)}{}_{\mu A} E_{(1)}{}^{\mu A} E_{(1)}{}_{\nu B} E_{(1)}{}^{\nu B}
-\frac{1}{2} c_{(11)}^2 E_{(1)}{}^{\mu A} E_{(1)}{}_{\mu}^{\ B} E_{(1)}{}^{\nu}_{\ A} E_{(1)}{}_{\nu B} \nn\\
&+& \frac{1}{2} c_{(22)}^2 E_{(2)}{}_{\mu A} E_{(2)}{}^{\mu A} E_{(2)}{}_{\nu B} E_{(2)}{}^{\nu B}
-\frac{1}{2} c_{(22)}^2 E_{(2)}{}^{\mu A} E_{(2)}{}_{\mu}^{\ B} E_{(2)}{}^{\nu}_{\ A} E_{(2)}{}_{\nu B} \nn\\
&+&  c_{(11)} c_{(12)} E_{(1)}{}_{\mu A} E_{(1)}{}^{\mu A} E_{(1)}{}^{\nu B} E_{(2)}{}_{\nu B}
- c_{(11)} c_{(12)} E_{(1)}{}^{\mu A} E_{(1)}{}_{\mu }^{\ B} E_{(1)}{}^{\nu}_{\ A} E_{(2)}{}_{\nu B} \nn\\
&+&  c_{(12)} c_{(22)} E_{(1)}{}^{\mu A } E_{(2)}{}_{\mu A } E_{(2)}{}_{\nu B } E_{(2)}{}^{\nu B }
- c_{(12)} c_{(22)} E_{(1)}{}^{\mu A } E_{(2)}{}_{\mu }^{\ B} E_{(2)}{}^{\nu }_{\ A } E_{(2)}{}_{\nu B} \nn\\
&-& \frac{1}{4} c_{(12)}^2 E_{(1)}{}^{\mu A }  E_{(1)}{}^{\nu }_{\ A } E_{(2)}{}_{\mu }^{\ B } E_{(2)}{}_{\nu B }
+ \frac{1}{4} c_{(12)}^2  E_{(1)}{}_{\mu A } E_{(1)}{}^{\mu A } E_{(2)}{}_{\nu B } E_{(2)}{}^{\nu B } \nn\\
&-& c_{(11)} c_{(22)} E_{(1)}{}^{\mu A } E_{(1)}{}^{\nu B } E_{(2)}{}_{\mu B } E_{(2)}{}_{\nu A } 
+ c_{(11)} c_{(22)} E_{(1)}{}^{\mu A } E_{(1)}{}^{\nu B } E_{(2)}{}_{\mu A }  E_{(2)}{}_{\nu B } \nn\\
&-& \frac{1}{4} c_{(12)}^2 E_{(1)}{}^{\mu A } E_{(1)}{}_{\mu }^{\ B} E_{(2)}{}^{\nu }_{\ A } E_{(2)}{}_{\nu B}.
\label{RH2D}
\eea
One can now compare \eqref{LH2D} and \eqref{RH2D} order-by-order so as to find the most generally allowed choice of $\alpha_{(ij)}$ that leads to consistent vielbein interactions. Comparing ${\cal O}(E_{(i)}^4)$ we find the matching condition $c_{(ii)} = \alpha_{(ii)}^2$, without any restrictions on the $\alpha$'s. Comparing ${\cal O}(E_{(i)}^3,E_{(j)}^1)$ we then find the matching condition $c_{(12)} = 2 \alpha_{(12)}$, yet again without any restrictions on the $\alpha$'s. This means we have used up all the freedom in the $c_{(ij)}$ -- all $c_{(ij)}$ coefficients have been fixed in terms of $\alpha_{(ij)}$ by matching ${\cal O}(E_{(i)}^4)$ and ${\cal O}(E_{(i)}^3,E_{(j)}^1)$ terms. But we have not yet matched ${\cal O}(E_{(i)}^2,E_{(j)}^2)$ terms. Doing so establishes the (necessary but not sufficient) requirement $\alpha_{(12)} = \alpha_{(11)} \alpha_{(22)}$ in order for the matter coupling to be ghost-free, i.e. it restricts the form of ${\bm g}^{(M)}$. In summary, we find the matching conditions
\begin{align}
&{\cal O}(E_{(i)}^4): &c_{(11)} &= \alpha_{(11)}^2  &c_{(22)} &= \alpha_{(22)}^2  \nn\\
&{\cal O}(E_{(i)}^3,E_{(j)}^1): &c_{(12)} &= 2 \alpha_{(12)}   \nn\\
&{\cal O}(E_{(i)}^2,E_{(j)}^2):  &\alpha_{(12)} &= \alpha_{(11)} \alpha_{(22)}.
\label{coeff-2D}
\end{align}
It is worth emphasising that, although which order we match first, which second etc. is arbitrary, the end result in terms of the $c_{(ij)}$ and $\alpha_{(ij)}$ parameters is always the same for a consistent matching. As a result we have constructed the general candidate matter coupling for Bigravity in 2D, which can be built out of vielbeins and the Lorentz metric, respects the weak equivalence principle and which does not introduce ghosts via effective cosmological constant contributions $\sqrt{-\det \left({\bm g}^{(M)}\right)}$. A coupling that satisfies these conditions, but takes on a different form than the one presented here will already introduce a ghost at or below the scale $\Lambda_3$. The relation of the $\alpha_{(ij)}$ to the $c_{(ij)}$ given in \eqref{coeff-2D} then uniquely specifies the associated consistent vielbein interaction terms generated by $\sqrt{-\det \left({\bm g}^{(M)}\right)}$. Substituting \eqref{coeff-2D} into \eqref{int-2D} we find
\bea
\sqrt{-\det \left({\bm g}^{(M)}\right)} \propto \tilde\epsilon_{A_1 A_2}\tilde\epsilon^{\mu_1\mu_2} &\Big(& \alpha_{(11)}^2  E_{(1)}{}_{\mu_1}^{\ A_1} E_{(1)}{}_{\mu_2}^{\ A_2} + 2 \alpha_{(11)}\alpha_{(22)} E_{(1)}{}_{\mu_1}^{\ A_1} E_{(2)}{}_{\mu_2}^{\ A_2} \nn\\
&+&\alpha_{(22)}^2 E_{(2)}{}_{\mu_1}^{\ A_1} E_{(2)}{}_{\mu_2}^{\ A_2}\Big).
\eea
Finally, collecting results \eqref{genmet2D} and \eqref{coeff-2D}, the general form of the matter metric for Bigravity in 2D is
\bea
{\bm g}^{(M)}_{\mn} &=&  \alpha_{(11)}^2 E^{(1)}{}_{\mu}^{\  A} E^{(1)}{}_{\nu}^{\ B} \eta_{AB} + \alpha_{(22)}^2 E^{(2)}{}_{\mu}^{\  A} E^{(2)}{}_{\nu}^{\ B} \eta_{AB} \nn\\
&+& \alpha_{(11)}\alpha_{(22)} \left( E^{(1)}{}_{\mu}^{\  A} E^{(2)}{}_{\nu}^{\ B} + E^{(2)}{}_{\mu}^{\  A} E^{(1)}{}_{\nu}^{\ B} \right) \eta_{AB},
\eea
so that we have found a two-parameter family of matter couplings, with the parameters being $\alpha_{(11)}$ and $\alpha_{(22)}$.

\subsection{Massive and Bigravity in 4D}\label{subsec-coupBi2D}

Having explicitly gone through the relatively straightforward 2D Bigravity case above, we now repeat the above procedure in order to establish a general consistent matter coupling in the context of Massive and Bigravity in 4D. In the Bigravity case consistent potential interactions for the two dynamical vielbeins $E_{(1)}$ and $E_{(2)}$ are governed by the action
\bea 
\label{drgtvielbi}
\cS_{\rm int} &=& - {m^2 M_{\rm eff}^{2}\over 8}\int \widehat\sum c_{(i_1 i_2 i_3 i_4)} {\bm \cI}_{(i_1 i_2 i_3 i_4)}\, ,
\eea
with $i_j \in \{ 1,2 \} $ and where the interaction terms ${\cI}$ are given in \eqref{2in4D}.
The generic matter metric following from \eqref{genmattermet} is still (just as in the 2D Bigravity case above)
\bea
{\bm g}^{(M)}_{\mn} &=& \alpha_{(11)}^2 E_{(1)}{}_{\mu}^{\  A} E_{(1)}{}_{\nu}^{\ B}\eta_{AB} + \alpha_{(22)}^2 E_{(2)}{}_{\mu}^{\  A} E_{(2)}{}_{\nu}^{\ B} \eta_{AB} \nn\\   &+& \alpha_{(12)} \left( E_{(1)}{}_{\mu}^{\  A} E_{(2)}{}_{\nu}^{\ B} + E_{(2)}{}_{\mu}^{\  A} E_{(1)}{}_{\nu}^{\ B}\right) \eta_{AB}
\eea
and in 4D we can express $\det \left({\bm g}^{(M)}\right)$ as 
\be
-\det \left({\bm g}^{(M)}\right) = -\frac{1}{4!}\tilde\epsilon^{\mu_1 \mu_2 \mu_3 \mu_4}\tilde\epsilon^{\nu_1 \nu_2 \nu_3 \nu_4} {\bm g}^{(M)}_{\mu_1 \nu_1}{\bm g}^{(M)}_{\mu_2 \nu_2} {\bm g}^{(M)}_{\mu_3 \nu_3}{\bm g}^{(M)}_{\mu_4 \nu_4}.
\ee
Proceeding as before by requiring
\be
\sqrt{-\det \left({\bm g}^{(M)}\right)} \propto \widehat\sum c_{(i_1 i_2 i_3 i_4)} {\cI}_{(i_1 i_2 i_3 i_4)},
\ee
we fix the coefficient of proportionality and explicitly solve this in terms of the $\alpha_{(ij)}$ and $c_{(i_1 i_2 i_3 i_4)}$) that satisfy
\be
-\det \left({\bm g}^{(M)}\right) = \frac{1}{(4!)^2}\left(\widehat\sum c_{(i_1 i_2 i_3 i_4)} {\cI}_{(i_1 i_2 i_3 i_4)}\right)^2.\label{detsol2}
\ee
Again matching terms order-by-order in \eqref{detsol2} we now find
\begin{align}
&{\cal O}(E_{(i)}^8): &c_{(1111)} &= \alpha_{(11)}^4  &c_{(2222)} &= \alpha_{(22)}^4  \nn\\
&{\cal O}(E_{(i)}^7,E_{(j)}^1): &c_{(1112)} &= 4 \alpha_{(11)}^2 \alpha_{(12)} &c_{(1222)} &= 4 \alpha_{(22)}^2 \alpha_{(12)}    \nn\\
&{\cal O}(E_{(i)}^6,E_{(j)}^2): &c_{(1122)} &= 6 \alpha_{(12)}^2  
&\alpha_{(12)} &= \alpha_{(11)} \alpha_{(22)}. \nn\\
\label{coeff-4Dbi}
\end{align}
Note that, as in the $2D$ case, matching the first two orders does not restrict the relation of the $\alpha$'s relative to one another at all. However, matching ${\cal O}(E_{(i)}^6,E_{(j)}^2)$ now already enforces $\alpha_{(12)} = \alpha_{(11)} \alpha_{(22)}$. The remaining orders not shown in \eqref{coeff-4Dbi} also match with the given choice of coefficients, confirming that this is indeed a consistent choice. Just as in $2D$, the $4D$ Bigravity metric is therefore restricted to take the form\footnote{As long as the symmetric vielbein condition is satisfied \cite{Deffayet:2012zc}, the form shown here is equivalent to the `acceptable' class of matter metrics proposed by \cite{deRham:2014naa}. Our approach shows that this is in fact the unique consistent way in which the matter metric can be formed out of vielbeins and the Lorentz metric $\eta_{AB}$ for Bigravity (where we require the absence of ghosts from the pure spin-2 piece generated by the matter action as before). We discuss the most general coupling to matter (i.e. in particular allowing inverse vielbeins in the construction of the matter metric as well) in the vielbein picture in \cite{mattercoupling}.}
\bea
{\bm g}^{(M)}_{\mn} &=&  \alpha_{(11)}^2 E^{(1)}{}_{\mu}^{\  A} E^{(1)}{}_{\nu}^{\ B} \eta_{AB} + \alpha_{(22)}^2 E^{(2)}{}_{\mu}^{\  A} E^{(2)}{}_{\nu}^{\ B} \eta_{AB} \nn\\
&+& \alpha_{(11)}\alpha_{(22)} \left( E^{(1)}{}_{\mu}^{\  A} E^{(2)}{}_{\nu}^{\ B} + E^{(2)}{}_{\mu}^{\  A} E^{(1)}{}_{\nu}^{\ B} \right) \eta_{AB}
\label{4Dbi-met}
\eea
In fact this form is independent of the dimension $D$, so this is the form of the Bigravity metric in any dimension. \eqref{coeff-4Dbi} also shows that the vielbein interactions generated by the matter action now satisfy
\bea \label{4Dbi-int}
\widehat\sum c_{(i_1 i_2 i_3 i_4)} {\cI}_{(i_1 i_2 i_3 i_4)} &=& \alpha_{(11)}^4 {\cI}_{(1111)} + \alpha_{(22)}^4 {\cI}_{(2222)} \\ &+& 4 \alpha_{(11)}^3 \alpha_{(22)} {\cI}_{(1112)}+ 4 \alpha_{(22)}^3 \alpha_{(11)} {\cI}_{(1222)}+ 6 \alpha_{(11)}^2 \alpha_{(22)}^2 {\cI}_{(1122)}\nn
\eea
up to an overall constant of proportionality.

Finally, the Massive Gravity potentials in the vielbein picture are simply special cases of the Bigravity interactions \eqref{drgtvielbi} with $E_{(2)} = \delta$. Essentially $E_{(2)}$ becomes the `vielbein' for the flat background metric $\eta$ and we can write ${\bf E}_{(2)} = {\bf 1} = \delta_\mu^{\ A} dx^\mu$. Substituting this into \eqref{4Dbi-met}, the matter metric becomes
\bea
{\bm g}^{(M)}_{\mn} &=&  \alpha_{(11)}^2 E^{(1)}{}_{\mu}^{\  A} E^{(1)}{}_{\nu}^{\ B} \eta_{AB} + \alpha_{(22)}^2  \eta_{\mn} \nn\\
&+& \alpha_{(11)}\alpha_{(22)} \left( E^{(1)}{}_{\mu}^{\  A} \delta_{\nu}^{\ B} + \delta_{\mu}^{\  A} E^{(1)}{}_{\nu}^{\ B} \right) \eta_{AB},
\eea
while the Massive Gravity interactions generated by the matter action take on the exact same form as in \eqref{4Dbi-int}. Note that $\eta_{\mn}$ can be straightforwardly replaced with any background reference metric \cite{Hassan:2011tf}, which we may want to couple to matter in addition to the dynamical metric  (in this case $\delta$ should be replaced by the `vielbein' for this non-dynamical background metric as well).

\subsection{Multi-Gravity in 4D}\label{subsec-coupMulti2D}

Finally we turn to the general case of a Multi-Gravity theory with $\N$ vielbeins in $D$ dimensions. We will display some of the intermediate steps for the specific case of a $4D$ spacetime, but just as before the form of the matter metric will not depend on $D$, so we will arrive at a general expression for acceptable matter metrics in the presence of $\N$ spin-2 fields.  We remind ourselves that our full matter metric takes on the form
\bea
{\bm g}^{(M)}_{\mn} &=& \sum_{i=1}^{\N} \alpha_{(ii)}^2 E^{(i)}{}_{\mu}^{\  A} E^{(i)}{}_{\nu}^{\ B} \eta_{AB} \nn\\
&+& \sum_{i,j=1,\ldots,\N}^{i < j} \alpha_{(ij)} \left( E^{(i)}{}_{\mu}^{\  A} E^{(j)}{}_{\nu}^{\ B} + E^{(j)}{}_{\mu}^{\  A} E^{(i)}{}_{\nu}^{\ B} \right) \eta_{AB},
\eea
where, as before, we wish to restrict the coefficients in this expressions by requiring that the pure vielbein interactions generated by this metric are ghost-free. We find solutions for the $\alpha_{(ij)}$ and $c_{(i_1 i_2 i_3 i_4)}$ that satisfy (in 4D)
\be \label{detMulti}
-\det \left({\bm g}^{(M)}\right) = \frac{1}{(4!)^2}\left(\widehat\sum c_{(i_1 i_2 i_3 i_4)} {\cI}_{(i_1 i_2 i_3 i_4)}\right)^2.
\ee
The explicit expressions for these terms are rather cumbersome, so we refrain from showing them here explicitly, but after some algebra we can solve \eqref{detMulti} order-by-order as before and find
\begin{align}
&{\cal O}(E_{(i)}^8): &c_{(iiii)} &= \alpha_{(ii)}^4    \nn\\
&{\cal O}(E_{(i)}^7,E_{(j)}^1): &c_{(iiij)} &= 4 \alpha_{(ii)}^2 \alpha_{(ij)} &c_{(ijjj)} &= 4 \alpha_{(jj)}^2 \alpha_{(ij)}    \nn\\
&{\cal O}(E_{(i)}^6,E_{(j)}^2): &c_{(iijj)} &= 6 \alpha_{(ij)}^2     &\alpha_{(ij)} &= \alpha_{(ii)} \alpha_{(jj)} \nn\\
&{\cal O}(E_{(i)}^6,E_{(j)}^1,E_{(k)}^1): &c_{(iijk)} &= 12 \alpha_{(ii)}^2 \alpha_{(jk)}    
&c_{(ijjk)} &= 12 \alpha_{(jj)}^2 \alpha_{(ik)} &c_{(ijkk)} &= 12 \alpha_{(kk)}^2 \alpha_{(ij)}      \nn\\
&{\cal O}(E_{(i)}^5,E_{(j)}^1,E_{(k)}^1,E_{(l)}^1): &c_{(ijkl)} &= 24 \alpha_{(ij)}\alpha_{(kl)}.\nn\\\label{coeff-Multi}
\end{align}
Several interesting points can be noted. Firstly, each order fixes the $c_{(i_1 i_2 i_3 i_4)}$ for a certain type of interaction: The first order fixes vielbein self interactions, the second order Bigravity interactions of the type $(iiij)$ and $(ijjj)$, the third order Bigravity interactions of the type $(iijj)$, the fourth order Tri-Gravity interactions and the fifth order interactions between four vielbeins. Also, the third order, as before, imposes the constraint $\alpha_{(ij)} = \alpha_{(ii)} \alpha_{(jj)}$, which is the only constraint relating only $\alpha$'s. Finally, note that there are another 10 different orders (all the other ways of partitioning the number 8 into four numbers, such as ${\cal O}(E_{(i)}^4,E_{(j)}^2,E_{(k)}^1,E_{(l)}^1)$ etc.), but no remaining free coefficients. However, with the mapping between between $c_{(i_1 i_2 i_3 i_4)}$ and $\alpha_{(j_1 j_2)}$ given by \eqref{coeff-Multi}, all orders match up, so that \eqref{detMulti} is always satisfied and the pure spin-2 interactions of the resulting matter action are consequently always ghost-free. 

The $c_{(i_1 i_2 i_3 i_4)}$ from \eqref{coeff-Multi} completely specify the vielbein interactions generated by the matter action in 4D and for the matter metric we now have the general expression (valid for $\N$ fields and any dimension $D$) 
\bea
{\bm g}^{(M)}_{\mn} &=& \sum_{i=1}^{\N} \alpha_{(ii)}^2 E^{(i)}{}_{\mu}^{\  A} E^{(i)}{}_{\nu}^{\ B} \eta_{AB} \nn\\
&+& \sum_{i,j=1,\ldots,\N}^{i < j} \alpha_{(ii)}\alpha_{(jj)} \left( E^{(i)}{}_{\mu}^{\  A} E^{(j)}{}_{\nu}^{\ B} + E^{(j)}{}_{\mu}^{\  A} E^{(i)}{}_{\nu}^{\ B} \right) \eta_{AB}
\label{mattermetmulti}
\eea
We have therefore constructed an $\N$-parameter family of matter metrics (and hence ways to couple $\N$ spin-2 fields to matter in the way outlined), where the parameters are $\alpha_{(11)}$ to  $\alpha_{(\N \N)}$. A particularly interesting consequence of this is related to the specific form of the vielbein interactions generated by the matter coupling (as specified via the $c_{(i_1 i_2 \ldots i_D)}$ in e.g. \eqref{coeff-Multi}). Coupling $\N$ spin-2 fields to matter in the way discussed will generate three- and four-vielbein interactions between all the participating spin-2 fields (for $\N \ge 4$). So, for example, one cannot consistently couple matter to three spin-2 fields in this way without introducing Tri- and Bigravity-like interactions between the spin-2 fields, though these interactions may of course be cancelled by adding appropriate interaction terms in the spin-2 self-interaction sector of the theory. But in the absence of such a `fine-tuning', coupling $\N > 2$ fields to matter and only retaining Bigravity-type interactions is not possible. This is a very different situation compared to the task of constructing consistent multi-gravity theories when one is only concerned with the spin-2 sector - there it is perfectly possible to build an $\N$ field theory only containing Bigravity interactions without any `fine-tuning' required \cite{Hinterbichler:2012cn,Noller:2013yja}.

\subsection{Matter vielbeins and the direct coupling to other matter fields}

We can rephrase our general result for the matter metric in a more concise form by considering the effective vielbein that matter couples to. The general matter metric \eqref{mattermetmulti}, satisfying all our requirements, can then be written 
\be \label{matter-viel-gen}
{\bm g}^{(M)}_{\mn} = E^{(M)}{}_{\mu}^{\  A} E^{(M)}{}_{\nu}^{\ B} \eta_{AB},
\ee
where we have introduced the effective `matter vielbein' $E^{(M)}$ associated to ${\bm g}^{(M)}_{\mn}$ and where, from \eqref{mattermetmulti}, $E^{(M)}$ satisfies
\be \label{vielbein-gen}
E^{(M)}{}_{\mu}^{\  A} = \sum_{i=1}^{\N} \alpha_{(ii)} E^{(i)}{}_{\mu}^{\  A}.
\ee
As such we can summarise our main result as saying that the general class of matter metrics in Massive, Bi- and Multi-Gravity, which are potentially ghost-free below $\Lambda_3$ and are constructed subject to imposing the weak equivalence principle and built out of vielbeins only, can be built from `matter vielbeins' , which are a general linear superposition of all vielbeins in the theory (i.e. all the vielbeins for the $\N$ spin-2 fields).
\footnote{We thank James Scargill for suggesting this way of formulating \eqref{mattermetmulti}.} The fact that we can write the matter coupling in terms of the matter vielbeins \eqref{vielbein-gen} is a consequence of the finding that $\alpha_{ij} = \alpha_{ii}\alpha_{jj}$ is required in order for the pure spin-2 interactions generated by the matter coupling to be consistent (see e.g. \eqref{coeff-Multi}). Note that, if $\alpha_{ij} = \alpha_{ii}\alpha_{jj}$ had not been required, of course the matter metric could still be written in terms of matter vielbeins as in \eqref{matter-viel-gen}, but the matter vielbeins would not take on the form \eqref{vielbein-gen}, but instead would be 'square-roots' of the full expression \eqref{genmattermet}. The non-trivial statement here is that the acceptable class of matter vielbeins takes on the very succinct form \eqref{vielbein-gen}. Any matter coupling that meets the above conditions, but is not built with matter vielbeins of the form \eqref{vielbein-gen} will generate a ghost below $\Lambda_3$. 

Finally we would like to reiterate that, for the class of metrics considered here, the form of \eqref{mattermetmulti} and \eqref{vielbein-gen} should be seen as a {\it necessary} condition for such `matter metrics' to lead to a ghost-free coupling to matter below $\Lambda_3$. A matter action constructed in accordance with the weak equivalence principle is always expected to give rise to an effective cosmological constant term for the `matter metric' as considered throughout this paper - the vacuum energy associated with the $\Phi_i$ will do so even in the absence of a bare cosmological constant in ${\cal L}\left[\Phi_i, {\bm g}^{(M)} \right]$ (also see \cite{deRham:2014naa}, where this is explicitly verified considering one-loop corrections in the case of Bigravity). Consequently the matter action will generate the pure spin-2 interactions considered here and it is imperative to ensure these are ghost-free. The explicit coupling to the matter fields $\Phi_i$ may, depending on the form of the matter action, introduce additional model-dependent restrictions on the general form of the matter metric \eqref{mattermetmulti} (in order to ensure ghost-freedom below some given energy scale $\Lambda$). We investigate the possibility of such further constraints from the direct coupling of spin-2 fields to the matter fields $\Phi_i$ in \cite{mattercoupling} - also see \cite{Yamashita:2014fga,deRham:2014naa}.

\section{Summary and Conclusions} \label{sec-conc}

In this paper we constructed a family of candidate couplings of matter to $\N$ spin-2 fields in $D$ dimensions in a way that does not inevitably introduce a Boulware-Deser ghost below $\Lambda_3$, assuming the weak equivalence principle. In particular, we found that it is indeed possible to couple matter to more than one spin-2 field, also in ways which retain the symmetry between different spin-2 fields present at the level of their self-interactions (also see appendix \ref{appendix} on this point). In our construction we coupled matter to general functions of the $\N$ vielbeins and the Lorentz metric $\eta_{AB}$ and, as a minimal criterion for consistency, required the pure spin-2 interactions generated by the matter action to be ghost-free. 
The resulting general family of ways in which matter can couple to `gravity' (the spin-2 sector) for $\N$ spin-2 fields in $D$ dimensions is given by
\be
\int d^D x \sqrt{-\det \left({\bm g}^{(M)}\right)} {\cal L}\left[\Phi_i, {\bm g}^{(M)} \right],
\ee
where the matter metric ${\bm g}^{(M)}$ satisfies (from \eqref{mattermetmulti}) 
\bea
{\bm g}^{(M)}_{\mn} &=& \sum_{i=1}^{\N} \alpha_{(ii)}^2 E^{(i)}{}_{\mu}^{\  A} E^{(i)}{}_{\nu}^{\ B} \eta_{AB} \nn\\
&+& \sum_{i,j=1,\ldots,\N}^{i < j} \alpha_{(ii)}\alpha_{(jj)} \left( E^{(i)}{}_{\mu}^{\  A} E^{(j)}{}_{\nu}^{\ B} + E^{(j)}{}_{\mu}^{\  A} E^{(i)}{}_{\nu}^{\ B} \right) \eta_{AB}
\label{conc-matmet}
\eea
and where we may express the `matter metric' ${\bm g}^{(M)}$ in terms of `matter vielbeins' $E^{(M)}$ satisfying the particular simple relation
\begin{align} \label{conc-viel}
{\bm g}^{(M)}_{\mn} &= E^{(M)}{}_{\mu}^{\  A} E^{(M)}{}_{\nu}^{\ B} \eta_{AB}, &E^{(M)}{}_{\mu}^{\  A} &= \sum_{i=1}^{\N} \alpha_{(ii)} E^{(i)}{}_{\mu}^{\  A}.
\end{align}

While this paper was being finished, \cite{Yamashita:2014fga,deRham:2014naa} appeared, which also discuss the coupling to matter in Bigravity (and consequently also in Massive Gravity). The Bigravity part of the classical analysis in this paper partially overlaps with that of \cite{deRham:2014naa} and where there is overlap our results agree. In fact, the result of \cite{deRham:2014naa} that the form of the effective metric in Bigravity is stable under quantum one-loop corrections\footnote{Note that, for a general effective matter metric, the mass scale $m$ associated with pure spin-2 interactions can however cease to be technically natural in these setups \cite{deRham:2014naa}.} is very encouraging, as this result should carry over to the vielbein Bigravity case considered here (at least as long as the symmetric vielbein condition is satisfied \cite{Deffayet:2012zc}) and we would expect it to apply to the general Multi-Gravity case as well, extending our classical analysis. Also note that, for the bigravity subset of matter couplings considered here, \cite{deRham:2014naa} have shown that in fact not only the pure spin-2 interactions, but also other interactions of the spin-2 field with matter do not re-excite the Boulware-Deser ghost up to the scale $\Lambda_3$ -- the full theory including explicit coupling to matter fields of the type discussed throughout here is therefore ghost-free up to $\Lambda_3$.

Several interesting topics for further research suggest themselves. The forms for the `matter metric' \eqref{conc-matmet} and `matter vielbein' \eqref{conc-viel} are {\it necessary} conditions for a ghost-free matter coupling -- the model-dependent precise coupling of all the matter degrees of freedom to the spin-2 fields may restrict this form further, i.e. they might impose additional restrictions on the $\alpha_{(ii)}$ in \eqref{conc-viel}. In \cite{mattercoupling} we extend and complete the argument presented here and discuss I) further restrictions resulting from this precise dependence of the matter action on the matter fields $\Phi_i$ (also see \cite{deRham:2014naa}) and II) the most general couplings to matter in the vielbein picture, where e.g. inverse vielbeins also partake in the construction of the effective matter metric. Investigating the effects on the background cosmology and the evolution of perturbations (particularly in the context of structure formation) in models with non-trivial matter couplings further should also prove interesting.
\\
    
\noindent {\bf Acknowledgements: } We would like to thank Tessa Baker, James Bonifacio, Claudia de Rham, Lavinia Heisenberg, Sigurd K. N\ae ss, Raquel Ribeiro  and especially Pedro Ferreira and James Scargill for many helpful discussions and comments on drafts of this paper. JN is supported by the STFC and BIPAC. The {\it xAct} package for Mathematica \cite{xAct} was used in the computation and check of some of the results presented here.

\appendix

\section{Appendix: A few concrete examples of consistent couplings to matter}   \label{appendix}

In this appendix we will spell out the allowed forms of the matter metric ${\bm g}^{(M)}$ for some concrete cases. If matter only couples to one metric only, ${\bm g}^{(M)} = g_{(i)}$, the only allowed coupling is the standard minimal one
\be
\int d^4x \sqrt{-\det \left({\bm g}^{(M)}\right)} {\cal L}\left[\Phi_i, {\bm g}^{(M)} \right] = \int d^4x \sqrt{-\det \left({\bm g}^{(i)}\right)} {\cal L}\left[\Phi_i, {\bm g}^{(i)} \right],
\ee
and the spin-2 interactions generated by the matter action are simply a (non-dynamical) cosmological constant term for $g_{(i)}$. In the presence of two spin-2 fields coupling to matter we have
\bea 
{\bm g}^{(M)}_{\mn} &=&  \alpha_{(11)}^2 E^{(1)}{}_{\mu}^{\  A} E^{(1)}{}_{\nu}^{\ B} \eta_{AB} + \alpha_{(22)}^2 E^{(2)}{}_{\mu}^{\  A} E^{(2)}{}_{\nu}^{\ B} \eta_{AB} \nn\\
&+& \alpha_{(11)}\alpha_{(22)} \left( E^{(1)}{}_{\mu}^{\  A} E^{(2)}{}_{\nu}^{\ B} + E^{(2)}{}_{\mu}^{\  A} E^{(1)}{}_{\nu}^{\ B} \right) \eta_{AB}.
\label{bimet-appendix}
\eea
If we set $\alpha_{(11)} = 0$ or $\alpha_{(22)} = 0$, matter minimally couples to only one of the two metrics again. Otherwise all terms in \eqref{bimet-appendix} are present (with any given numerical values that the $\alpha_{(ii)}$ take). In particular notice the case where $\alpha_{(11)} = \alpha_{(22)}$. Then the coupling to matter retains the symmetry between $E_{(1)}$ and $E_{(2)}$ that the self-interaction part of the Bigravity action possesses. In this case the matter metric takes on the form
\bea 
{\bm g}^{(M)}_{\mn} &\propto &  E^{(1)}{}_{\mu}^{\  A} E^{(1)}{}_{\nu}^{\ B} \eta_{AB} +  E^{(2)}{}_{\mu}^{\  A} E^{(2)}{}_{\nu}^{\ B} \eta_{AB} \nn\\
&+& \left( E^{(1)}{}_{\mu}^{\  A} E^{(2)}{}_{\nu}^{\ B} + E^{(2)}{}_{\mu}^{\  A} E^{(1)}{}_{\nu}^{\ B} \right) \eta_{AB},
\label{symmetric-bimet-appendix}
\eea
where $\alpha_{(11)}^2 = \alpha_{(22)}^2 $ determines the constant of proportionality. In this case the `matter vielbein' $E^{(M)}$ satisfies
\begin{align} \label{sym-viel}
E^{(M)}{}_{\mu}^{\  A} &= \alpha_{(11)} \sum_{i=1}^{\N} E^{(i)}{}_{\mu}^{\  A}.
\end{align}

For the general matter metric for coupling to $\N$ spin-2 fields
\bea
{\bm g}^{(M)}_{\mn} &=& \sum_{i=1}^{\N} \alpha_{(ii)}^2 E^{(i)}{}_{\mu}^{\  A} E^{(i)}{}_{\nu}^{\ B} \eta_{AB} \nn\\
&+& \sum_{i,j=1,\ldots,\N}^{i < j} \alpha_{(ii)}\alpha_{(jj)} \left( E^{(i)}{}_{\mu}^{\  A} E^{(j)}{}_{\nu}^{\ B} + E^{(j)}{}_{\mu}^{\  A} E^{(i)}{}_{\nu}^{\ B} \right) \eta_{AB},
\label{mattermetmulti-appedix}
\eea
matter can again either just minimally couple to a single spin-2 field via $g_{(i)}$ (in which case $\alpha_{(jj)} = 0$ for all $j \neq i$). Alternatively matter couples to two spin-2 fields in which case the matter coupling generates all possible Bigravity interaction terms in the way described, or matter couples to $\N$ spin-2 fields and the matter coupling generates all possible $\N$-gravity interaction terms in $D$ dimensions (for $\N > D$ these will be all the ways of building $D$-Gravity with the different ways of choosing $D$ out of $\N$ spin-2 fields).  Symmetry between all the $\N$ spin-2 fields can be enforced by requiring   $\alpha_{(ii)} = \alpha_{(jj)}$ for all $i,j$ as before and \eqref{sym-viel} will still correctly identify the corresponding symmetric `matter vielbein' $E^{(M)}$.

\bibliographystyle{JHEP}
\bibliography{mcb-bib}

\providecommand{\href}[2]{#2}\begingroup\raggedright\begin{thebibliography}{10}

\bibitem{deRham:2010ik}
C.~de~Rham and G.~Gabadadze, {\it {Generalization of the Fierz-Pauli Action}},
  {\em Phys.Rev.} {\bf D82} (2010) 044020,
  [\href{http://xxx.lanl.gov/abs/1007.0443}{{\tt arXiv:1007.0443}}].

\bibitem{deRham:2010kj}
C.~de~Rham, G.~Gabadadze, and A.~J. Tolley, {\it {Resummation of Massive
  Gravity}},  {\em Phys.Rev.Lett.} {\bf 106} (2011) 231101,
  [\href{http://xxx.lanl.gov/abs/1011.1232}{{\tt arXiv:1011.1232}}].

\bibitem{Hassan:2011hr}
S.~Hassan and R.~A. Rosen, {\it {Resolving the Ghost Problem in non-Linear
  Massive Gravity}},  {\em Phys.Rev.Lett.} {\bf 108} (2012) 041101,
  [\href{http://xxx.lanl.gov/abs/1106.3344}{{\tt arXiv:1106.3344}}].

\bibitem{Hassan:2011tf}
S.~Hassan, R.~A. Rosen, and A.~Schmidt-May, {\it {Ghost-free Massive Gravity
  with a General Reference Metric}},  {\em JHEP} {\bf 1202} (2012) 026,
  [\href{http://xxx.lanl.gov/abs/1109.3230}{{\tt arXiv:1109.3230}}].

\bibitem{Hassan:2011zd}
S.~Hassan and R.~A. Rosen, {\it {Bimetric Gravity from Ghost-free Massive
  Gravity}},  {\em JHEP} {\bf 1202} (2012) 126,
  [\href{http://xxx.lanl.gov/abs/1109.3515}{{\tt arXiv:1109.3515}}].

\bibitem{Hassan:2011ea}
S.~Hassan and R.~A. Rosen, {\it {Confirmation of the Secondary Constraint and
  Absence of Ghost in Massive Gravity and Bimetric Gravity}},  {\em JHEP} {\bf
  1204} (2012) 123, [\href{http://xxx.lanl.gov/abs/1111.2070}{{\tt
  arXiv:1111.2070}}].

\bibitem{Hinterbichler:2012cn}
K.~Hinterbichler and R.~A. Rosen, {\it {Interacting Spin-2 Fields}},  {\em
  JHEP} {\bf 1207} (2012) 047, [\href{http://xxx.lanl.gov/abs/1203.5783}{{\tt
  arXiv:1203.5783}}].

\bibitem{Hinterbichler:2011tt}
K.~Hinterbichler, {\it {Theoretical Aspects of Massive Gravity}},  {\em
  Rev.Mod.Phys.} {\bf 84} (2012) 671--710,
  [\href{http://xxx.lanl.gov/abs/1105.3735}{{\tt arXiv:1105.3735}}].

\bibitem{deRham:2014zqa}
C.~de~Rham, {\it {Massive Gravity}},
  \href{http://xxx.lanl.gov/abs/1401.4173}{{\tt arXiv:1401.4173}}.

\bibitem{Tamanini:2013xia}
N.~Tamanini, E.~N. Saridakis, and T.~S. Koivisto, {\it {The Cosmology of
  Interacting Spin-2 Fields}},  {\em JCAP} {\bf 1402} (2014) 015,
  [\href{http://xxx.lanl.gov/abs/1307.5984}{{\tt arXiv:1307.5984}}].

\bibitem{Akrami:2013ffa}
Y.~Akrami, T.~S. Koivisto, D.~F. Mota, and M.~Sandstad, {\it {Bimetric gravity
  doubly coupled to matter: theory and cosmological implications}},  {\em JCAP}
  {\bf 1310} (2013) 046, [\href{http://xxx.lanl.gov/abs/1306.0004}{{\tt
  arXiv:1306.0004}}].

\bibitem{Akrami:2014lja}
Y.~Akrami, T.~S. Koivisto, and A.~R. Solomon, {\it {The nature of spacetime in
  bigravity: two metrics or none?}},
  \href{http://xxx.lanl.gov/abs/1404.0006}{{\tt arXiv:1404.0006}}.

\bibitem{Yamashita:2014fga}
Y.~Yamashita, A.~De~Felice, and T.~Tanaka, {\it {Appearance of Boulware-Deser
  ghost in bigravity with doubly coupled matter}},
  \href{http://xxx.lanl.gov/abs/1408.0487}{{\tt arXiv:1408.0487}}.

\bibitem{deRham:2014naa}
C.~de~Rham, L.~Heisenberg, and R.~H. Ribeiro, {\it {On couplings to matter in
  massive (bi-)gravity}},  \href{http://xxx.lanl.gov/abs/1408.1678}{{\tt
  arXiv:1408.1678}}.

\bibitem{Nibbelink:2006sz}
S.~Groot~Nibbelink, M.~Peloso, and M.~Sexton, {\it {Nonlinear Properties of
  Vielbein Massive Gravity}},  {\em Eur.Phys.J.} {\bf C51} (2007) 741--752,
  [\href{http://xxx.lanl.gov/abs/hep-th/0610169}{{\tt hep-th/0610169}}].

\bibitem{Hanada:2008hs}
T.~Hanada, K.~Shinoda, and K.~Shiraishi, {\it {Multi-graviton theory in
  vierbein formalism}},  \href{http://xxx.lanl.gov/abs/0801.2641}{{\tt
  arXiv:0801.2641}}.

\bibitem{Chamseddine:2011mu}
A.~H. Chamseddine and V.~Mukhanov, {\it {Massive Gravity Simplified: A
  Quadratic Action}},  {\em JHEP} {\bf 1108} (2011) 091,
  [\href{http://xxx.lanl.gov/abs/1106.5868}{{\tt arXiv:1106.5868}}].

\bibitem{Chamseddine:2011bu}
A.~H. Chamseddine and M.~S. Volkov, {\it {Cosmological solutions with massive
  gravitons}},  {\em Phys.Lett.} {\bf B704} (2011) 652--654,
  [\href{http://xxx.lanl.gov/abs/1107.5504}{{\tt arXiv:1107.5504}}].

\bibitem{Mirbabayi:2011aa}
M.~Mirbabayi, {\it {A Proof Of Ghost Freedom In de Rham-Gabadadze-Tolley
  Massive Gravity}},  {\em Phys.Rev.} {\bf D86} (2012) 084006,
  [\href{http://xxx.lanl.gov/abs/1112.1435}{{\tt arXiv:1112.1435}}].

\bibitem{Deffayet:2012zc}
C.~Deffayet, J.~Mourad, and G.~Zahariade, {\it {A note on 'symmetric' vielbeins
  in bimetric, massive, perturbative and non perturbative gravities}},  {\em
  JHEP} {\bf 1303} (2013) 086, [\href{http://xxx.lanl.gov/abs/1208.4493}{{\tt
  arXiv:1208.4493}}].

\bibitem{Banados:2013fda}
M.~Ba\~nados, C.~Deffayet, and M.~Pino, {\it {The Boulware-Deser mode in 3D
  first-order massive gravity}},  {\em Phys.Rev.} {\bf D88} (2013), no.~12
  124016, [\href{http://xxx.lanl.gov/abs/1310.3249}{{\tt arXiv:1310.3249}}].

\bibitem{Mourad:2014roa}
J.~Mourad and D.~Steer, {\it {Translation invariant time-dependent solutions to
  massive gravity II}},  {\em JCAP} {\bf 1406} (2014) 058,
  [\href{http://xxx.lanl.gov/abs/1405.1862}{{\tt arXiv:1405.1862}}].

\bibitem{Deser:2014hga}
S.~Deser, M.~Sandora, A.~Waldron, and G.~Zahariade, {\it {Covariant constraints
  for generic massive gravity and analysis of its characteristics}},
  \href{http://xxx.lanl.gov/abs/1408.0561}{{\tt arXiv:1408.0561}}.

\bibitem{Hassan:2012wt}
S.~Hassan, A.~Schmidt-May, and M.~von Strauss, {\it {Metric Formulation of
  Ghost-Free Multivielbein Theory}},
  \href{http://xxx.lanl.gov/abs/1204.5202}{{\tt arXiv:1204.5202}}.

\bibitem{Noller:2013yja}
J.~Noller, J.~H.~C. Scargill, and P.~G. Ferreira, {\it {Interacting spin-2
  fields in the St\"uckelberg picture}},  {\em JCAP} {\bf 1402} (2014) 007,
  [\href{http://xxx.lanl.gov/abs/1311.7009}{{\tt arXiv:1311.7009}}].

\bibitem{loops}
J.~H.~C. Scargill, J.~Noller, and P.~G. Ferreira, {\it {Cycles of interactions
  in multi-gravity theories}},  \href{http://xxx.lanl.gov/abs/1410.7774}{{\tt
  arXiv:1410.7774}}.

\bibitem{Comelli:2012vz}
D.~Comelli, M.~Crisostomi, F.~Nesti, and L.~Pilo, {\it {Degrees of Freedom in
  Massive Gravity}},  {\em Phys.Rev.} {\bf D86} (2012) 101502,
  [\href{http://xxx.lanl.gov/abs/1204.1027}{{\tt arXiv:1204.1027}}].

\bibitem{ASM-thesis}
A.~Schmidt-May, {\it {Thesis, to be published}}, .

\bibitem{cutoff}
J.~Bonifacio and J.~Noller, {\it {On strong coupling scales in massive
  gravity}},  \href{http://xxx.lanl.gov/abs/1412.4780}{{\tt arXiv:1412.4780}}.

\bibitem{Folkerts:2011ev}
S.~Folkerts, A.~Pritzel, and N.~Wintergerst, {\it {On ghosts in theories of
  self-interacting massive spin-2 particles}},
  \href{http://xxx.lanl.gov/abs/1107.3157}{{\tt arXiv:1107.3157}}.

\bibitem{Hinterbichler:2013eza}
K.~Hinterbichler, {\it {Ghost-Free Derivative Interactions for a Massive
  Graviton}},  \href{http://xxx.lanl.gov/abs/1305.7227}{{\tt arXiv:1305.7227}}.

\bibitem{Kimura:2013ika}
R.~Kimura and D.~Yamauchi, {\it {Derivative interactions in de
  Rham-Gabadadze-Tolley massive gravity}},  {\em Phys.Rev.} {\bf D88} (2013)
  084025, [\href{http://xxx.lanl.gov/abs/1308.0523}{{\tt arXiv:1308.0523}}].

\bibitem{deRham:2013tfa}
C.~de~Rham, A.~Matas, and A.~J. Tolley, {\it {New Kinetic Interactions for
  Massive Gravity?}},  {\em Class.Quant.Grav.} {\bf 31} (2014) 165004,
  [\href{http://xxx.lanl.gov/abs/1311.6485}{{\tt arXiv:1311.6485}}].

\bibitem{Copeland:2006wr}
E.~J. Copeland, M.~Sami, and S.~Tsujikawa, {\it {Dynamics of dark energy}},
  {\em Int.J.Mod.Phys.} {\bf D15} (2006) 1753--1936,
  [\href{http://xxx.lanl.gov/abs/hep-th/0603057}{{\tt hep-th/0603057}}].

\bibitem{Clifton:2011jh}
T.~Clifton, P.~G. Ferreira, A.~Padilla, and C.~Skordis, {\it {Modified Gravity
  and Cosmology}},  {\em Phys.Rept.} {\bf 513} (2012) 1--189,
  [\href{http://xxx.lanl.gov/abs/1106.2476}{{\tt arXiv:1106.2476}}].

\bibitem{mattercoupling}
S.~Melville and J.~Noller {\em {in preparation.}}

\bibitem{Boulware:1973my}
D.~Boulware and S.~Deser, {\it {Can gravitation have a finite range?}},  {\em
  Phys.Rev.} {\bf D6} (1972) 3368--3382.

\bibitem{xAct}
J.~M. Mart\'in-Garc\'ia, {\it {xAct 2002-2014}},  {\em http://www.xact.es/}.

\end{thebibliography}\endgroup

\end{document}